\documentclass[runningheads]{llncs}
\usepackage[T1]{fontenc}
\usepackage{graphicx}
\usepackage{longtable}

\PassOptionsToPackage{hyphens}{url}\usepackage{hyperref}
\usepackage{xcolor}

\begin{document}
\title{Intelligent Urban Traffic Management via Semantic Interoperability across Multiple Heterogeneous Mobility Data Sources}
%\title{Addressing Interoperability Issues for Mobility Data Sources through Semantic Web Technologies}
%
\titlerunning{Intelligent Urban Traffic Management via Semantic Interoperability}

\author{Mario Scrocca\inst{1}\orcidID{0000-0002-8235-7331} \and
Marco Grassi\inst{1}\orcidID{000-0003-3139-3049} \and
Marco Comerio\inst{1}\orcidID{0000-0003-3494-9516} \and
Valentina Anita Carriero\inst{1}\orcidID{0000-0003-1427-3723} \and
Tiago Delgado Dias\inst{2}\orcidID{0009-0009-1785-7119}\and 
Ana Vieira Da Silva\inst{2}\orcidID{0000-0003-0328-1787}
\and
Irene Celino\inst{1}\orcidID{0000-0001-9962-7193}
}
\authorrunning{M. Scrocca et al.}
\institute{Cefriel -- Politecnico di Milano \\Viale Sarca 226, 20126 Milano, Italy
 \\ \email{\emph{name.surname}@cefriel.com} \and
 A-to-Be Mobility Technology, S.A. \\ Edificio Brisa 2785-599 São Domingos de Rana, Portugal \\ \email{\{tiago.delgado.dias,ana.vieira.silva\}@a-to-be.com}
 }
\maketitle              % typeset the header of the contribution
\begin{abstract}
The integrated exploitation of data sources in the mobility domain is key to providing added-value services to passengers, transport companies and authorities. Indeed, multiple stakeholders operate and maintain different kinds of data but several interoperability issues limit their effective usage. In this paper, we present an architecture enabled by Semantic Web technologies to overcome such issues and facilitate the development of an integrated solution for mobility stakeholders. The proposed solution is composed of different components that address challenges for enabling data interoperability, from the findability of data sources to their integrated consumption adopting standardised data formats. We report on the implementation and validation in four European cities of the TANGENT solution enabling data-driven tools for the dynamic management of multimodal traffic. Finally, we discuss the feedback received by users testing the solution and the lessons learnt during its development.

\end{abstract}

\section{Introduction}\label{sec:introduction}

Interoperability is one of the main challenges to enable collaboration between travel and transport industry players. The interoperability of data and services is essential for creating an ecosystem of transport stakeholders enabling the definition of new data-driven solutions. The benefits range from enhancing the operations of transport companies to the provision of integrated and seamless mobility services to users.
The TANGENT project\footnote{\url{https://doi.org/10.3030/955273}}, co-funded by the European Commission under the Horizon 2020 Programme, is developing new complementary ICT tools for optimising traffic operations in a coordinated and dynamic way from a multimodal perspective.  In this paper, we discuss how knowledge graphs and semantic technologies can help in tackling interoperability in the mobility domain.

The development and testing of the data-driven solutions developed by TANGENT in the Athens, Lisbon, Greater Manchester, and Rennes Metropole case studies asked for a solution to different data interoperability challenges. The definition of a proper solution for data sharing and usage is not straightforward due to several issues to be addressed: datasets in different formats and/or using different data models, data services relying on different specifications and technologies, and metadata describing data sources according to different profiles. 

This paper describes the design and implementation of an integrated set of tools aimed at addressing data interoperability issues for heterogeneous data sources from different stakeholders by employing Semantic Web technologies. 
%The proposed solution consists of four components: (i) a digital platform enabling data-sharing and findability while enforcing governance; (ii) a reference conceptual model defining common semantics; (iii) a set of semantic harmonisation and fusion pipelines to fulfil integration requirements associated with data sources; (iv) a uniform mechanism to access data sources. 
While the proposed solution can be applied to a generic domain, we describe how we implemented and tested it considering the specificities of multimodal transportation. Furthermore, we describe the reusable resources, like metadata specification and ontologies, that are made publicly available. 

Within the integrated TANGENT solution, data interoperability is leveraged to support the deployment of an innovative dynamic traffic management platform. The platform provides intelligent services and cutting-edge user interfaces and is enabled in each city by the integrated consumption of several data sources. The data sources were retrieved and harmonised involving different stakeholders and systems, thus demonstrating the flexibility and scalability of the proposed solution. We received positive feedback from users during testing sessions that acknowledged the benefits of integrating data for different transportation modes within a single solution and expressed their interest in adopting the solution within their operations. 
%were interested in expanding the scope of the tested platform
%and they will evaluate its acquisition as a commercial product after the completion of the project.

The remainder of the paper is organised as follows. Section~\ref{sec:challenges} discusses the motivating challenges and the related work. Section~\ref{sec:solution} describes the proposed solution and its implementation considering the heterogeneous mobility data sources of the four cities involved in the validation. Section~\ref{sec:validation} discusses the evaluation and lessons learned. Finally, Section~\ref{sec:conclusions} draws the conclusions.

\section{Challenges and related work}\label{sec:challenges}

Data interoperability is a challenging objective to enable different stakeholders to communicate and exchange information effectively without losing meaning. Indeed, stakeholders adopt different (legacy) systems for data management and exchange that cannot be directly integrated or harmonised. Five major challenges can be identified and should be addressed~\cite{Comerio2022DataReq}:
%
%\begin{enumerate}
%\item
  1. \textbf{Locate} (\emph{which data is available and where?}),
%\item
  2. \textbf{Access} (\emph{how to obtain the needed data?}),
%\item
  3. \textbf{Harmonise} (\emph{how to convert data according to the required
  data model?}),
%\item
  4. \textbf{Integrate} (\emph{how to ensure different data sources can be
  merged?}),
%\item
  5. \textbf{Extract} (\emph{how to consume harmonized and integrated
  data?}).
%\end{enumerate}

Each of these challenges is associated with several issues and identifying a single solution is impossible since a single interoperability problem cannot be formulated. Indeed, data interoperability scenarios are widely heterogeneous and pose various requirements~\cite{Carenini2020SPRINT} that can be possibly faced only by
considering a set of tools appropriately configured. To select such
tools and define an integrated solution~\cite{tra2024harmonisation}, we reviewed state-of-the-art data interoperability solutions based on Semantic Web technologies and their application for the mobility domain.

\subsection{Locate and access}\label{locate-and-access}

The first challenge is the findability and discoverability of data. Data cannot be re-used and (made) interoperable if they cannot be found. For this reason, data catalogues/portals are implemented %providing a solution 
to describe data sources %, i.e., datasets and data services, 
through a set of metadata. The challenge is associated with the need for a proper, structured and machine-readable description of data sources that could also support interoperability across different data catalogues. Once data sources are located, the second challenge is related to data accessibility. Data catalogues adopt different strategies for data access mainly associated with the architectural choices for the hosting and storage of static and dynamic data sources. The challenge is to enable uniform access to heterogeneous data sources for end users.

The \emph{locate} and \emph{access} challenges are also being addressed by the European Commission through National Access Points (NAP) for mobility data. Each Member State should operate a NAP to enable the sharing of mobility data by transport stakeholders as mandated by dedicated Delegated Regulations~\cite{Commission2013Delegated,Commission2015Delegated,Commission2017Delegated} supplementing the ITS Directive 2010/40/EU~\cite{Parliament2010Directive}. The concept of NAP leverages the one of Data Catalogue, i.e., a digital platform to facilitate the sharing of data sources and their findability by other stakeholders. However, several mobility data platforms exist but are not interoperable. %without effective interoperability solutions. 
Even in the case of NAPs, each Member State adopted different approaches for their implementation, thus creating interoperability issues at the European level. For this reason, the NAPCORE\footnote{\url{https://napcore.eu/}} project is currently working on coordinating and harmonizing such platforms around Europe. One important objective is  supporting the findability of data contained in each mobility data platform~\cite{Scrocca2022Towards} and defining mobilityDCAT-AP\footnote{\url{https://w3id.org/mobilitydcat-ap}}, a uniform metadata specification to access the data sources. The adoption of structured metadata descriptors according to well-known vocabularies (e.g., DCAT-AP~\cite{DCAT-AP}) is fundamental to facilitate search % the implementation of searching functionalities 
within one or multiple data catalogues. Moreover, proper data governance must be defined to regulate the usage of the catalogue between the different involved stakeholders. Finally, data catalogues should support the harmonisation of technological  access to data sources. % and/or provide detailed documentation to assist the data user.

For these reasons, we identified the two core components of the proposed solution as a shared \emph{Data Catalogue} to enable the findability of data sources and a uniform \emph{Data API} for accessibility.

\subsection{Harmonise, Integrate, and Extract}\label{sec:hie}

The remaining three data interoperability challenges (harmonise,
integrate, and extract) are related to the processing of (meta)data to enable their integration and exploitation according to common semantics. A flexible solution is required to address heterogeneous requirements in terms of:
%
%\begin{itemize}
%\item
  (a) \textbf{schema and data transformation}: information manipulation %of the encoded   information 
  to obtain syntactic (structural) and semantic
  interoperability of (meta)data;
%\item
  (b) \textbf{integration} with existing information systems as data sources (i.e.,
  components generating or storing the data) and/or data sinks (i.e.,
  components consuming or archiving the data).
%\end{itemize}

Different approaches can be exploited and implemented, spanning from ad-hoc solutions targeting a specific scenario to more general and scalable
solutions supporting multiple stakeholders and data representations. The semantic any-to-one mapping approach based on~\cite{Vetere2005Models} and validated in~\cite{Scrocca2020Turning}
reduces the number of mappings, i.e., translations from one
representation to another, that are needed to implement interoperability by
different stakeholders. Such an approach is based on the identification of a reference model for the domain of interest. Each stakeholder is responsible for defining mappings from their own data representation to the reference model (\emph{lifting}) and vice versa (\emph{lowering}). 
%The main advantage is the definition of an interoperable and integrated RDF graph, modelled according to a formal reference ontology, as an additional valuable product of the conversion procedure. 
In this paper, we discuss how we adopted this approach. % can be integrated into an overall data architecture to address broader interoperability issues and how an intermediate knowledge graph can effectively address harmonisation and fusion requirements for heterogeneous datasets and data services.

Considering the mobility domain, different reference models are proposed based on existing standards. Chouette~\cite{Gendre2011CHOUETTE} and the SNAP solution~\cite{ruckhaus_applying_2023} rely on a reference
model based on  Transmodel\footnote{\url{https://www.transmodel-cen.eu/}} for the conversion of
Public Transport (PT) data in different formats. The \texttt{transit\_model} tool\footnote{\url{https://github.com/hove-io/transit\_model}}
adopts the Navitia Transit Feed Specification (NTFS)\footnote{\url{https://github.com/hove-io/ntfs-specification}}
to manage, convert and enrich transit data from/to different formats.
Moreover, considering traffic data, the Datex II\footnote{\url{https://www.datex2.eu/}}
specification is often used as a reference model to convert custom data
formats and share harmonised data~\cite{Guerreiro2016architecture}.

Different approaches for lifting and lowering can be suitable considering a specific scenario. Moreover, the harmonisation, integration and extraction process may require the definition of custom pre- and post-processing, considering different interoperability issues. Therefore, composing and configuring different components should be possible considering the specific requirements for integrating certain data sources.

Different semantic-based ETL (Extract, Transform and Load) tools have been proposed %in the literature to address the problem of 
to define composable procedures with Semantic Web technologies~\cite{Grassi2023Composable}. Technologies for declarative knowledge graph construction~\cite{VanAssche2022Declarative} can effectively support lifting transformations, while a standardised lowering solution to convert RDF to any format using a generic declarative language is currently missing~\cite{scrocca2024not}. Moreover, other components, such as message filtering or routing, are usually required within a transformation pipeline. Enterprise Integration Patterns~\cite{Hohpe2004Enterprise} offer a relevant categorisation of the components and techniques for system integration.

%UnifiedViews~\cite{Knap2014UnifiedViews:} and LinkedPipes~\cite{Klímek2016LinkedPipes} provide environments to feed and curate RDF knowledge bases. A different approach is used by Talend4SW\footnote{\url{https://github.com/fbelleau/talend4sw}}, aiming to complement an already existing tool (Talend) with the components required to interact with RDF data.

In conclusion, two additional components are identified to support the solution: a \emph{Reference Conceptual Model} defining common semantics and the composition and configuration of \emph{Semantic Harmonisation and Fusion Pipelines}.

\section{TANGENT solution for dynamic and intelligent multimodal traffic management}\label{sec:solution}

\begin{figure}[t!]
    \centering
    \includegraphics[width=\linewidth]{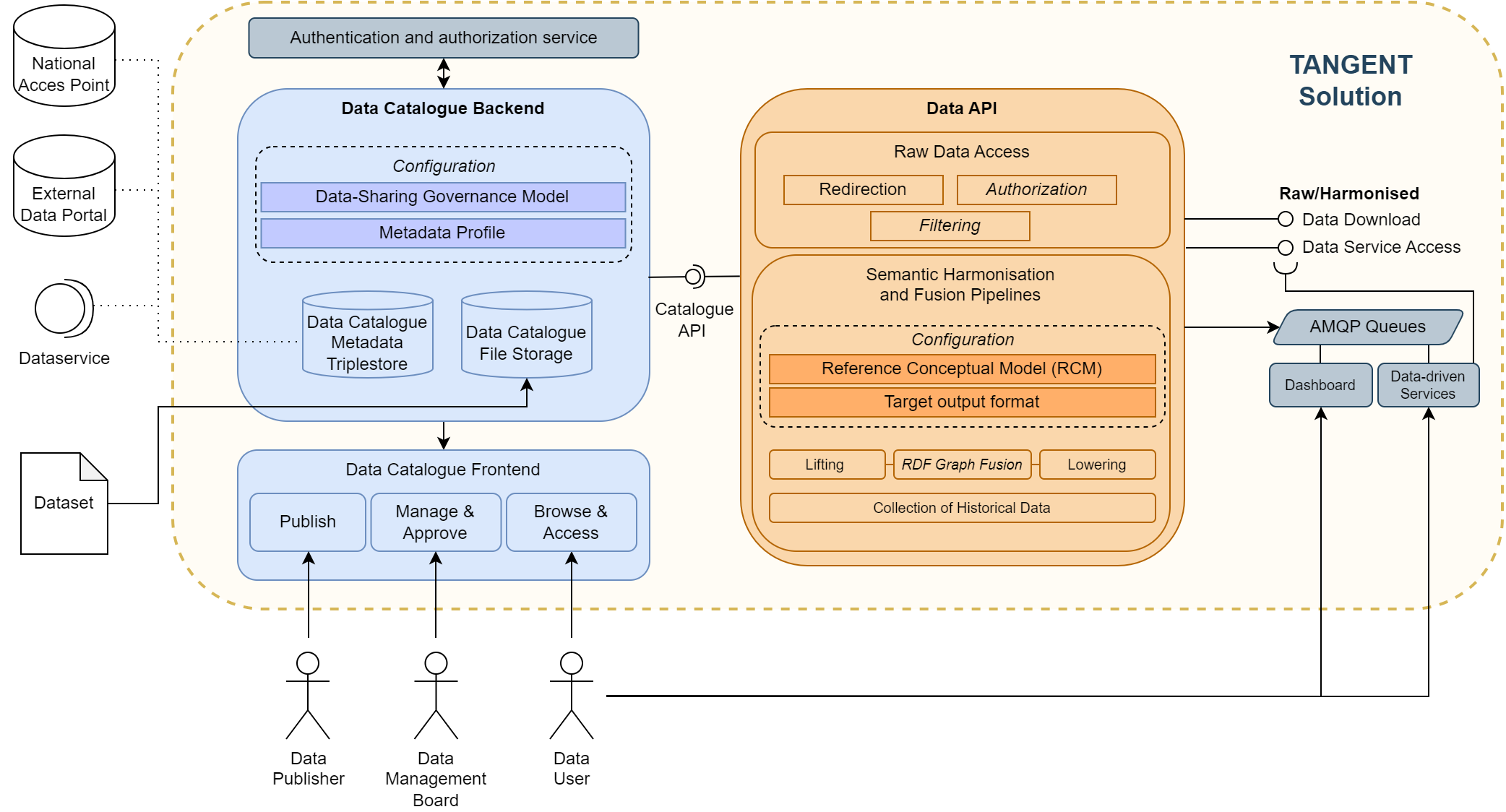}
    \caption{Overview of the proposed solution for data interoperability.}
    \label{fig:architecture}
\end{figure}

This section describes the TANGENT solution for dynamic and intelligent multimodal traffic management. We discuss each macro-component of the proposed architecture and then its integration within the overall TANGENT solution~\cite{tra2024architecture} as shown in Figure~\ref{fig:architecture}. The main original contributions of this work are the implementation and integration of different technologies to propose an holistic architecture for data interoperability based on Semantic Web technologies and its customisation for the multimodal traffic management domain. In the following we describe them, highlighting their value in solving the discussed challenges and their impact on the business scenarios.

\subsection{Data Catalogue}\label{sec:data-catalogue}

This component is a catalogue of digital assets available online and accessible by users via a web browser. Different digital assets can be characterized by specifying a metadata descriptor according to a common metadata profile. %Metadata defines a set of information that a user interacting with the catalogue can insert/visualize for an asset of that type. 
%The data catalogue may act both as a metadata platform, i.e., storing descriptions of data sources hosted elsewhere, or as a data platform, i.e., actually storing certain data (e.g., due to confidentiality issues). 
The data catalogue may also harvest metadata descriptions from existing data portals (e.g., NAPs). Programmatic access to the list of assets published and their metadata should be enabled via a dedicated API (\emph{Catalogue API}). Metadata serialized in RDF enable advanced functionalities based on querying and/or automated processing by agents of such metadata (e.g., in federation scenarios). Finally, the catalogue  enforces processes for the governance of digital assets. 
%The governance model should define key processes for data-management to be addressed by stakeholders with different roles and following specific rules. 

The  \emph{Data Catalogue} provides a single location where all the collected data sources for each city are described and can be explored. Its development is based on the Knowledge Catalog and Governance (KCONG) framework\footnote{\url{https://kcong.cefriel.com/}}, developed by Cefriel, which is customised considering the TANGENT \emph{Data Sharing Governance Model}~\cite{tra2024governance,Comerio2022} and the TangentDCAT-AP metadata specification as metadata profile.
%, that implements the characteristics described in Section~\ref{sec:architecture} for the Data Catalogue component. 

%\subsubsection{Data-Sharing Governance model}
%\textbf{Data-Sharing Governance model}
The  \emph{Data-Shar\-ing Governance model} supports the strategic and operational management of data-sharing. The model focuses on the tasks/proc\-ess\-es related to providing access to data sources needed by the various technical components. It identifies key processes (data publication, data quality, data access, data storage, data usage) to be addressed by stakeholders with different roles and following specific rules. The identified roles are: (i) \emph{Data publisher}, a person responsible for publishing and describing a data source within the catalogue; (ii) \emph{Data Management Board (TMB)}, a group of people responsible for the management and control of a (set of) data source(s) within the catalogue; (iii) \emph{Data user}: a person accessing and using a data source available in the catalogue.
%Each TANGENT case study leader acts as a TANGENT Data Publisher for data sources related to their case study. 
As an example of the defined rules (fully described in~\cite{Comerio2022}), the Lisbon case study leader acts as a data publisher and can create and modify only the metadata descriptions of data sources related to the Lisbon case study. %However, the Lisbon case study leader can also visualize the metadata descriptions of data sources for the other case studies but cannot modify them.

%\subsubsection{TangentDCAT-AP metadata profile}
%\textbf{TangentDCAT-AP metadata profile}
The definition of the \emph{TangentDCAT-AP} metadata specification considered best practices, particularly the reuse of well-known vocabularies for metadata. For this reason, TangentDCAT-AP is defined as an extension of the DCAT Application Profile~\cite{DCAT-AP}, considering the requirements for mobility data platforms elicited by the NAPCORE project~\cite{Scrocca2022Towards} and specific requirements for the description of data sources elicited within the TANGENT project~\cite{Comerio2022}. The final release of TangentDCAT-AP is compatible with the first official release of mobilityDCAT-AP\footnote{\url{https://w3id.org/mobilitydcat-ap/releases/1.0.0/}} by the NAPCORE project. mobilityDCAT-AP represents an extension of DCAT-AP focusing on requirements for data sources in the mobility domain and will be officially recommended as the reference metadata specification for National Access Points\footnote{\url{https://napcore.eu/release-of-the-mobilitydcat-ap/}} and adopted for the European Mobility Data Space\footnote{\url{https://www.linkedin.com/company/deployemds}}. %The documentation of 
TangentDCAT-AP is available at \url{https://knowledge.c-innovationhub.com/tangent/tandcatap}. The additional properties defined by TangentDCAT-AP have been also published online together with the defined controlled vocabularies following best practices~\cite{Scrocca2022Towards}. The published vocabularies are hosted on GitHub\footnote{\url{https://github.com/cefriel/tandcatap}} and served through content negotiation. The vocabularies define possible statuses assigned to a data source, data requirements to categorise the content of data sources and their types.

\begin{figure}[t!]
    \centering
    \includegraphics[width=\linewidth]{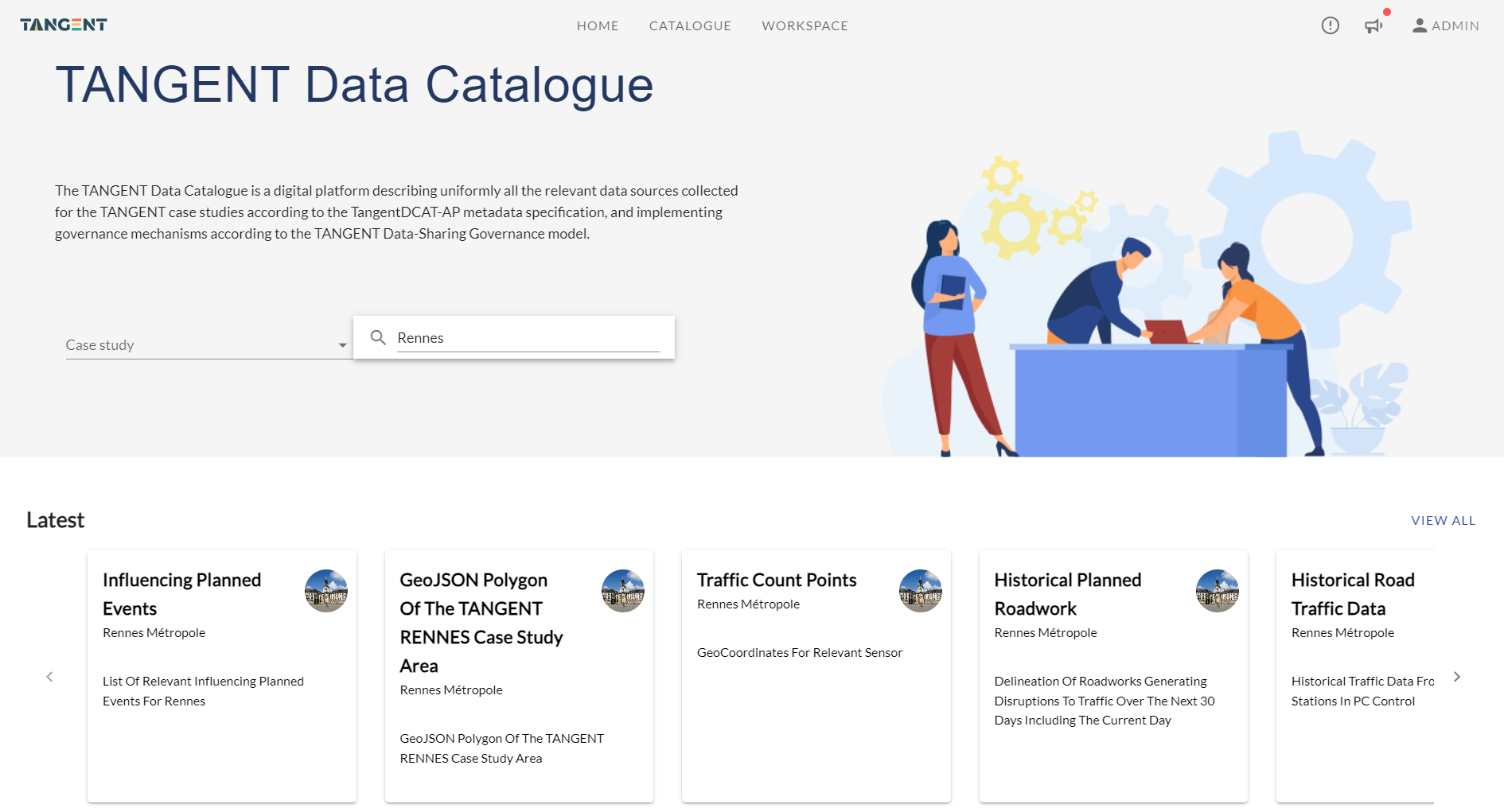}
    \caption{Homepage of the Data Catalogue.}
    \label{fig:catalogue}
\end{figure}

Figure~\ref{fig:catalogue} shows the homepage of the Data Catalogue and some available data sources. Three main areas are available to the user: the homepage, the catalogue and the workspace. The homepage is designed to show the latest changes and 
%offer the shortest possible path to one of t
the most common interactions with the catalogue, i.e. searching. The catalogue area allows the user to browse all available data sources and perform more fine-grained searches.  Lastly, the workspace area allows visualising/editing a specific data source, or getting an overview of the state of the data sources owned by the current user. An external identity and access management solution is used for authentication and authorization of different users to access data sources.

Currently, the Data Catalogue contains metadata about 145 data sources.

\subsection{Reference Conceptual Model}\label{rcm}

The \emph{Reference Conceptual Model} (RCM) supports the representation of heterogeneous information from different data sources through a common ontological model to enable shared semantics and interoperability. The model is based on existing data standards to adopt the correct domain terminology and covers the representation of all the entities and properties required to implement meaningful data exchanges among the involved stakeholders.

In the mobility domain, several ontological models have been proposed. However, they cover specific requirements, and it is not possible to identify a generic and well-adopted ontology~\cite{Katsumi2018Ontologies}. For these reasons, we started our work by an\-a\-lys\-ing existing data standards to support the identification of the relevant semantics. Then, following the best practice of reusing existing models~\cite{DBLP:series/ssw/GarijoP20}, relevant ontologies to be included in the RCM were identified. 

We started with the analysis of data standards requested by the European Commission (EC) Delegated Regulations (DR) mentioned in Section~\ref{locate-and-access}. 
The data standards mentioned in those directives are:
%\begin{itemize}
%\item
  (i) \emph{DATEX II} (\url{https://www.datex2.eu/}): the EU standard for the exchange of traffic-related data; %(traffic situations, traffic status, traffic management, Messages displayed on Variable Message Signs (VMS), service information, parking, truck parking, urban traffic specifics, electromobility infrastructure, refuelling and recharging, management of electronic traffic regulations, and urban vehicle access regulations);
%\item
  (ii) \emph{NeTEx} (\url{https://netex-cen.eu/}): the CEN Technical Standards for exchanging Public Transport schedules and related data; %(public transport network topologies, public transport scheduled timetables, and public transport fares);
%\item
  (iii) \emph{SIRI} (\url{https://www.siri-cen.eu/}): the CEN technical standard for the exchange of real-time information about the planned, current, or projected performance of public transport operations. %(real-time information about schedules, vehicles, and éconnections, together with general informational messages related to the operation of the services).
%\end{itemize}

To support the definition of the RCM, the existing ontologies encoding the semantics of the mentioned standards have been analysed. 

Considering the DATEX II format, two different models have been identified. The first one, directly developed by the DATEX II organization% and presented in 2021 at the DATEX II 6th Forum 
\cite{2021DATEX}, is a JSON-LD serialisation of the DATEX II conceptual model version 3, divided into five main modules\footnote{Models used available at \url{https://datex2.eu/vocab/3}. Additional modules are now available to accommodate the new versions of the DATEX II specification.}% reflecting the DATEX II specification modules
: {Payload}, \emph{Common}, \emph{LocationReferencing}, \emph{Situation}, \emph{Road Traffic Data}, \emph{Variable Message Sign}.
The second model~\cite{Gutierrez2022LOD-RoadTran18} based on DATEX II was developed by the LOD-RoadTran18
%\footnote{\url{https://cef.uv.es/lodroadtran18}} 
project to support the publication of DATEX II data as Linked Open Data. 
%An endpoint exposing Czech National Access Point data as Linked Open Data is available online\footnote{\url{https://registr.dopravniinfo.cz/en/sources/cz-ndic\_lod-srti-sparql/}}. 
The advantage of the first model is its full coverage with respect to the DATEX II specification, the one-to-one mapping to classes and properties, and the fact that it is directly defined and published by the DATEX II organization; however, this model is defined as an almost automatic conversion of the DATEX II specification.
%resulting in poor design decision from an ontological point of view. 
The advantage of the second model is that the LOD-RoadTran18 project followed proper ontology engineering methodologies to define the model; however, 
%the LOD SRTI DATEX II ontology\footnote{\url{https://cef.uv.es/lodroadtran18/def/transporte/dtx\_srti}} 
it covers only a portion of the DATEX II specification. For these reasons, we selected the first model for the RCM.
%As already defined in the first release of the model, the DATEX II JSON-LD ontology is selected as a basis to represent the traffic-related information since it guarantees complete coverage of the DATEX II specification.

The NeTEx and SIRI standards are based on the Transmodel\footnote{\url{https://www.transmodel-cen.eu/standards/}} conceptual model. The Mobility Ontology Catalogue\footnote{\url{https://w3id.org/mobility}} defines a suite of ontologies based on existing standards, including a Transmodel ontology. The Transmodel ontology, firstly defined within the SNAP\footnote{\url{https://snap-project.eu/}} project, has been extended and reviewed over the years~\cite{ruckhaus_applying_2023} and currently defines five submodules: \emph{Core}, \emph{Commons}, \emph{Fares}, \emph{Facilities}, \emph{Journeys}. The Transmodel ontology does not cover the entire Transmodel but was used to effectively support mappings to NeTEx~\cite{Scrocca2020Turning}. 
%The Linked GTFS ontology~\cite{2023Linked} is an alternative ontology for static public transportation data directly based on the GTFS specification, thus not aligned with the Transmodel. 
To the best of our knowledge, a dedicated SIRI ontology does not exist. However, since SIRI is derived from Transmodel, the current Transmodel ontology can be exploited to represent common concepts and possibly extended to represent the missing ones. We developed and published a dedicated ontology representing concepts and relations mapped from SIRI, %and considered as relevant to TANGENT's requirements. 
%For the design of the ontology, we 
adopting an approach aligned to the one leveraged for the definition of the DATEX II JSON-LD ontology from the DATEX II specification. %Therefore, we did not follow a classic ontology engineering process starting from scratch, but we adapted the existing SIRI specification to an ontological format. 
Moreover, we manually curated the SIRI ontology to improve the alignment with the other ontologies adopted in the RCM. 
%We implemented the ontology using the Web Ontology Language 2 (OWL)\footnote{\url{https://www.w3.org/TR/owl-rdf-based-semantics/}} and we published it online fulfilling the best practices for ontology implementation and publication. 
The SIRI ontology is available and documented at \url{https://knowledge.c-innovationhub.com/siri}.% (preferred prefix \emph{siri:})
%\footnote{A content negotiation mechanism is available to provide access to both human-readable documentation and machine-readable serialisations in the RDF format. The GitHub repository \url{https://www.github.com/cefriel/siri} is used to host and maintain the ontology}. 
The current version of the ontology (v1.0.0) 
%does not cover the entire SIRI specification but 
focuses on the modelling of 
%the fulfilment of TANGENT data requirements by considering 
situations affecting the transport network and monitored vehicle journeys.
%A draft GTFS-RT (\url{https://w3id.org/mobility/gtfs-rt}) ontology is available in the GitHub repository\footnote{\url{https://github.com/oeg-upm/mobility/tree/main/gtfs-rt}} of the Mobility Ontology Catalogue.

Based on the performed analysis, the RCM was defined as a suite of ontologies considering the semantics of relevant EU-mandated standards and the already available related ontologies. The DCI Metadata Terms
%\footnote{\url{http://purl.org/dc/terms/}} 
and Schema.org
%\footnote{\url{https://schema.org/}} 
vocabularies are reused by the Transmodel ontology and similarly also in other RCM modules.
Table~\ref{tab:rcm} provides a complete overview of the ten different modules defined for the  \emph{Reference Conceptual Model}, summarising the considered base standard  for  concepts and relationships, the data requirements covered by the module, and the reused ontologies. 

\begin{table}[t]
\centering
\resizebox{\textwidth}{!}{%
\begin{tabular}{l|c|l|l}
\multicolumn{1}{c|}{\textbf{Module}} & \textbf{Base Standard} & \multicolumn{1}{c|}{\textbf{Data   Requirements}} & \multicolumn{1}{c}{\textbf{Reused   ontologies}} \\ \hline \hline
Road Transport   Network & Datex II & \begin{tabular}[c]{@{}l@{}}Road   Transport Network \\ (roads, limited access zones, etc)\end{tabular} & \begin{tabular}[c]{@{}l@{}}GeoSPARQL,   \\ Basic Geo, \\ Datex II JSON-LD (location, common)\end{tabular} \\ \hline
Road Equipment & Datex II & Road Equipment Position & \begin{tabular}[c]{@{}l@{}}Datex II JSON-LD (location), \\ DC Terms, Schema.org\end{tabular} \\ \hline
Road Traffic   Data & Datex II & \begin{tabular}[c]{@{}l@{}}Road Traffic Measurements \\ (traffic occupancy, speed, flow)\\ Floating Vehicle Data (GPS, mobile, etc)\end{tabular} & Datex II JSON-LD (traffic) \\ \hline
Road Travel   Times & Datex II & \begin{tabular}[c]{@{}l@{}}Road Travel Times (external services, \\ statistics, etc.)\end{tabular} & Datex II JSON-LD (location, traffic) \\ \hline
Events & Datex II & \begin{tabular}[c]{@{}l@{}}Road Transport Network \\ Events (planned) / Incidents (unplanned)\\ Influencing Planned \\ Events (sports, entertainment, etc)\\ Weather  Events\end{tabular} & \begin{tabular}[c]{@{}l@{}}Datex II JSON-LD (situation, \\ location, common)\end{tabular} \\ \hline
Weather Data & Datex II & \begin{tabular}[c]{@{}l@{}}Forecasted Weather Data\\ Weather Data (measurements, \\ e.g., temperature, humidity, etc.)\end{tabular} & \begin{tabular}[c]{@{}l@{}}Datex II JSON-LD (location, traffic, \\ common)\end{tabular} \\ \hline
Stop Points & NeTEx & Public Transport Network & \begin{tabular}[c]{@{}l@{}}Transmodel   ontology (commons, \\ journeys), Basic Geo\end{tabular} \\ \hline
Schedules & NeTEx & \begin{tabular}[c]{@{}l@{}}Public Transport Schedules \\ and Lines\end{tabular} & \begin{tabular}[c]{@{}l@{}}Transmodel ontology (commons,   \\ journeys, organisations)\end{tabular} \\ \hline
Situation   Exchange & SIRI & \begin{tabular}[c]{@{}l@{}}Public Transport Network \\ Events (planned) / Incidents (unplanned)\end{tabular} & Basic Geo \\ \hline
Vehicle Monitoring & SIRI & \begin{tabular}[c]{@{}l@{}}Floating PT Vehicle Data\\ Public Transport Delays\end{tabular} & Basic Geo \
\end{tabular}%
}
\vspace{0.1cm}
\caption{Overview of the TANGENT Reference Conceptual Model}
\label{tab:rcm}
\vspace{-2em}
\end{table}

The latest release of the RCM is published online\footnote{\url{https://knowledge.c-innovationhub.com/tangent/schema}} including the documentation of the different modules and the serialisation of the additionally defined classes/properties. %\footnote{The best practices mentioned for the SIRI ontology are also followed in this case and the GitHub repository \url{https://www.github.com/cefriel/tangent-model} is used for hosting and maintenance}.
%The TANGENT Reference Conceptual Model is complemented by the definition of custom classes and properties covering gaps in the existing ontologies and specific requirements of the considered use case.
The definition of the RCM has been guided by the requirements elicited in TANGENT for the harmonisation and fusion of data sources for multimodal traffic management. %Indeed, we did not aim to define a fully comprehensive suite of ontologies to describe the entire mobility domain but to identify and extend the ones needed to cover the project requirements. 
Nevertheless, the RCM is made available and we recommend its reuse and extension to address additional requirements. 

\subsection{Semantic Harmonisation and Fusion Pipelines}\label{sec:semantic-harmonisation-and-fusion-pipelines}

Transformation pipelines should support the harmonisation and fusion of data sources available by leveraging the \emph{Reference Conceptual Model}. The definition of such pipelines requires the elicitation of harmonisation and fusion requirements considering the analysis of raw data sources collected in the \emph{Data Catalogue}, and the definition of the information required by the other components that are integrated into the solution and the related target output format. We mainly addressed the requirements of two downstream data usage: 
the need for large-scale historical data for training of machine learning models to support traffic management, the need to access static and (quasi) real-time data to empower a set of innovative applications for traffic managers.
%, which need access to data at runtime (e.g., to provide visualizations to users). 
In both cases, the need to overcome data heterogeneity and sparsity requires transformation pipelines.
The basic pipeline is composed of a lifting operation, a (set of) graph operations (e.g., to perform data fusion), and a lowering operation. %A particular case of such pipelines can be implemented for the collection and aggregation of data from real-time data sources to generate historical datasets that are relevant to train machine learning models.

A flexible and scalable technology for the implementation of the pipelines should provide (i) a set of reusable building blocks that can be configured according to specific requirements, and (ii) a declarative approach to configure the lifting and lowering transformations without developing ad-hoc and hard-to-maintain solutions.
%The TANGENT Semantic Harmonisation and Fusion Pipelines were configured to address the harmonisation and fusion requirements elicited by technical partners, considering the data sources collected in the Data Catalogue and their expected usage to enable the overall TANGENT solution. As discussed in Section~\ref{sec:hie}, the proposed semantic harmonisation and fusion process requires a lifting and a lowering step. The target output of the lifting process is RDF data aligned with the defined TANGENT Reference Conceptual Model. The result of the lowering step depends instead on the requirements of the data user. 
% Most of the pipelines target the interoperability of data exchanged with the overall TANGENT solution. Additional pipelines are defined to support the harmonisation and fusion of data from real-time data sources to generate historical data sources.
%Multiple TANGENT data requirements can be associated with the same module of the TANGENT Reference Conceptual Model and, consequently, with the same JSON Schema. Considering the defined mappings (i) a single data source can be harmonised considering multiple target JSON Schemas, (ii) data sources associated with different data requirements can be mapped to the same target JSON Schema, and (iii) a reduced set of API Data Requirements for the exchange of harmonised data can be identified for WP6 in comparison to the ones originally presented in D6.4.
Chimera\footnote{\url{https://github.com/cefriel/chimera}}
\cite{Grassi2023Composable} is an open-source solution based on Apache Camel to enable the definition of semantic data transformation pipelines with different components for knowledge graph construction, transformation, validation, and exploitation.
%and was tested in the SPRINT project also to support ORM-based approaches~\cite{Carenini2021SPRINT}. 
The advantage of Chimera is its integration
with Apache Camel, providing off-the-shelf %out-of-the-box 
and production-ready components to implement
Enterprise Integration Patterns and % a wide set of production-ready components 
to integrate pipelines with heterogeneous
systems (e.g., HTTP API, WebSocket, MQTT). 
%oreover, custom components can be easily defined and integrated within a pipeline (e.g., to implement custom extraction/filtering approaches). 
For these reasons, Chimera was selected to implement the semantic any-to-one mapping approach and Apache Camel is leveraged to implement the Data API and smoothly integrate the defined pipelines.

%The Chimera\footnote{\url{https://github.com/cefriel/chimera}} framework  has been leveraged to define and implement the TANGENT Semantic Harmonisation and Fusion Pipelines. Chimera is a Java-based software solution that leverages the Apache Camel integration framework toachieve a low-code approach for defining the pipelines. Indeed, the framework provides a set of ready-to-use building blocks that can be configured together with the Apache Camel Components to address heterogeneous scenarios for data harmonisation/fusion. Different approaches can be used to configure a pipeline, i.e., an Apache Camel \emph{Route}, exploiting Domain Specific Languages\footnote{\url{https://camel.apache.org/manual/dsl.html}}. The Chimera components are made available through Maven Central\footnote{https://search.maven.org/artifact/com.cefriel/chimera} to facilitate their adoption in Apache Camel projects. On the one hand, a Chimera pipeline can leverage Apache Camel components\footnote{\url{https://camel.apache.org/components/3.20.x/index.html}} for data acquisition and delivery tasks through well-established protocols, e.g., HTTP API, WebSocket, MQTT. On the other hand, Chimera provides a set of components to work with RDF Graphs and perform different operations for Graph Construction, Graph Transformation, Graph Validation, and Graph Exploitation.

%Chimera currently supports RML-based lifting (RML Component) and template-based lifting and lowering based on Apache Velocity (Mapping Template Component).
The data transformation from a source format and source semantics to a target ontology, i.e., the lifting process, can be handled by either the RML Component or the Mapping Template Component of Chimera. The lifted data, in the form of an RDF graph aligned with the Reference Conceptual Model, can be manipulated using the operations defined by the Chimera Graph Component. For example, these operations enable the fusion of data and/or the filtering/extraction of certain information. The Mapping Template Component handles the lowering process from RDF to the target data format\footnote{Regarding the lowering, we initially investigated the possibility of applying JSON-LD frames~\cite{json-ld-framing} to convert the RDF Graph to the target JSON Schemas. However, we encountered difficulties in addressing cases in which the structure of the RDF graph does not directly correspond to the structure of the target JSON.}.
%Each implemented pipeline combines several Apache Camel and Chimera components in a Camel \emph{Route}. The basic harmonisation pipeline
%consists in a lifting step from the raw data format to an RDF representation considering the target module of the TANGENT Reference Conceptual Model and a lowering step from RDF to the corresponding JSON Schema. 
For the TANGENT pipelines, we decided to use the Mapping Template component since it can be applied for both lifting and lowering~\cite{scrocca2024not}.
%provides more flexibility in the definition of complex mapping rules with respect to a full declarative approach like RML~\cite{scrocca2024not}. Moreover, it can be applied both for lifting and lowering, and could be more easily adopted by developers not familiar with RDF, like those who were involved in the implementation. The data and schema transformations executed in each pipeline are configured through a set of templates defined using the Mapping Template Language (MTL) that is documented online\footnote{Examples and Wiki at \url{https://github.com/cefriel/mapping-template}}.

The definition of harmonisation and fusion requirements for the solution required the analysis of (i) raw data sources collected in the Data Catalogue, and (ii) the information required by the other components to be integrated into the overall solution. As a result of a collaborative effort among partners in charge of developing downstream data-driven services, we identified the requirements in terms of target output format and the set of data sources to be harmonised.
Considering the target output, we identified the need for a harmonised CSV format to support the training of traffic prediction machine learning models, and of JSON schemas to feed the real-time services at runtime via AMQP\footnote{\url{https://www.amqp.org/}} queues. The semantics of the RCM was used as a basis to support both the annotation of columns in CSV and of fields in JSON\footnote{JSON Schemas are available at \url{https://github.com/cefriel/tangent-model/json-schemas} and contain a simplified representation of the information modelled in RDF to minimise the amount of exchanged data; if the data should be consumed as JSON-LD, a proper context can be associated with each JSON message.}.

% A set of JSON Schemas were collaboratively defined to support the semantics of the RCM and the need for JSON data flows defined for the overall TANGENT architecture. We defined a JSON Schema
%\footnote{https://json-schema.org/} 
%for each module in the RCM\footnote{Available online at \url{https://github.com/cefriel/tangent-model/json-schemas}}. Moreover,

All in all, we were able to harmonize and fuse data from 43 data sources adopting different data formats (mainly CSV, XML, JSON) and more than 30 data models across the 4 urban case studies. Indeed, many data sources were based on custom data models, thus requiring dedicated lifting mappings, and we could reuse them only for GTFS\footnote{\url{https://developers.google.com/transit/gtfs/}} feeds and data from the same data provider. On the other hand, we could leverage the any-to-one approach to define a single lowering mapping for each target output (10 lowering mappings to JSON schemas). Additionally, via dedicated pipelines, we generated 8 historical datasets targeting a CSV format. These pipelines are based on the same lifting mappings but are configured to regularly (e.g., 1-minute frequency) collect, harmonise and fuse data from real-time data sources. We run these pipelines for over 6 months collecting 92GB of compressed data.
%by defining x lifting mappings, y lower mappings and z pipelines.

\subsection{Data API}\label{sec:data-api}

The \emph{Data API}  provides uniform access to data sources collected through the Data Catalogue and represents the integration point for the overall integrated solution. The Data API aims at solving access issues for both \emph{raw data sources} (raw data as collected and shared by the data publisher) and \emph{harmonised and/or fused data sources} (data produced as the result of a semantic harmonisation and fusion pipeline). Harmonised data sources are represented in the Data Catalogue as different \emph{Distribution}s\footnote{\url{https://www.w3.org/ns/dcat\#Distribution}} of the same data source, i.e., different serialisations of the same information. The same approach is also used for data sources provided in multiple raw formats, e.g., CSV and JSON. The result of a fusion process is instead added to the catalogue as a new record since it represents a new data source.
The Data API gives access to two main types of data sources: (i) \emph{datasets} usually directly downloadable from a specific URL, and (ii) \emph{data services}  implementing different interaction mechanisms (e.g., a REST API). % and should be adequately invoked to access the data. 
The Data API implements the API Gateway pattern~\cite{Microservices}, thus providing a single and coherent entry point for the final user. The user can access a data source by knowing the endpoint at which the Data API is located and the identifier of the data source to be accessed. If the user is authorized through the Data Catalogue, the Data API handles authorization mechanisms for the different data sources in a transparent way and provides access to them. Moreover, in cases where a data source should be filtered according to specific requirements, the Data API can be configured to provide access only to the relevant data (e.g., adding the proper parameters to filter data according to a defined temporal/geographical scope). A dedicated parameter can be used to request a specific distribution of the data source (e.g., the harmonised format).
The Data API was implemented using the Apache Camel
framework since: (i) it provides all the relevant components
to interact with the different data platforms (NAPs, Open data/private portals, etc.) hosting data
sources, (ii) it can be easily integrated with the semantic
harmonisation and fusion pipelines. 

The integration of an external data source published on the Data Catalogue within the Data API requires different steps. First of all, we perform an analysis of the accessibility metadata provided for the Dataset (access URL/download URL) or the Data service (endpoint URL/endpoint documentation). If the data source is not available online, we contact the
responsible stakeholder  to get access. We leverage the storage layer of the Data Catalogue  to upload the data and we evaluate the need to define a data service.
Then, we investigate the expected data access interaction (e.g., used protocol); in particular, we evaluate authorization and authentication mechanisms. We configure the required Apache Camel component to retrieve the data source (e.g., HTTP component), and define the integration logic (e.g, filtering the data of a data source considering the relevant temporal/spatial scope for the case study).
Finally, if available, we integrate the semantic harmonisation and fusion pipeline  to allow users to request data in their harmonised format.
Once a data source is integrated into the Data API, the corresponding metadata to access it are updated in the Data Catalogue. In total, we developed 87 integrations to provide access via the Data API to all the raw/harmonised data sources approved for usage.

\section{Evaluation and Lessons Learned}\label{sec:validation}

This section discusses an evaluation of the proposed solution considering various perspectives and the lessons learned.

% technical evaluation: qui potresti "dare i numeri" per spiegare la scala della soluzione che non è un toy example; inoltre potresti spiegare che la parte di raccolta e armonizzazione dei dati storici fa guadagnare tempo infinito ai data scientist che vogliono addestrare modelli di machine learning (e anche avere un modello di riferimento permette di spiegare quali dati NON ci sono?) [Novelty in the application or assessment of Semantic Web and Knowledge Graph technologies, which can be reflected in terms of, for example: (1) the role they play in the solution; (2) how they foster adoption; or (3) their combination/interplay with other technologies.]
%\subsubsection{Technical evaluation} 
\textbf{Technical evaluation} The integrated on-cloud deployment consisted of testing and production environments with a cluster of four virtual machines and a set of managed services (e.g., database and load balancer). The runtime data for the four case studies are handled by 32 AMQP message exchanges fed by the Data API. The current data are refreshed with different periodicity depending on the considered data source and are organised with static data in 91 distinct collections on the database. The solution processes 130 messages per minute at peak time and manages around 4 GB of data at a time only to visualise the network's current status. The implemented solution demonstrates the feasibility and advantages of adopting an approach for data harmonisation and fusion based on Semantic Web technologies also to support real-time visualisations and data-driven services in a production-ready environment. The semantic any-to-one approach supported good scalability considering the high number of data sources involved and their substantial heterogeneity. Moreover, we demonstrated how it is possible to define system integrations that seamlessly combine semantic harmonisation and fusion to enable interoperability of data exchanges. We also positively assessed that the executed transformations introduced negligible latency (in the order of milliseconds) considering the update frequency of the data sources (often in the order of minutes). 
Finally, we highlighted the possibility of leveraging the same solution to generate datasets for training machine learning models. A defined pipeline for harmonisation was easily configured to collect and aggregate harmonised data from real-time data sources. Such an approach, relying on common semantics, reduces enormously the effort needed by data scientists to assess heterogeneous input datasets and facilitates the reuse of training algorithms (e.g., for different cities considering different data sources).
%\item user evaluation: qui quasi quasi accorperei le due valutazioni del data catalogue e della dashboard, dicendo che è stata fatta una valutazione qualitativa con n utenti di business e riportando solo i commenti che coprono la nostra parte (chissene della visualizzazione o delle città con dati diversi), sottolineando findability e accessibility di dati eterogenei [    Evidence of the adoption of the proposed solution by a relevant user base (domain practitioners, the general public, developers, etc.), preferably distinct from the proposer’s institutions and the Semantic Web and Knowledge Graph research communities.]

\begin{figure}[t!]
    \centering
    \includegraphics[width=\linewidth]{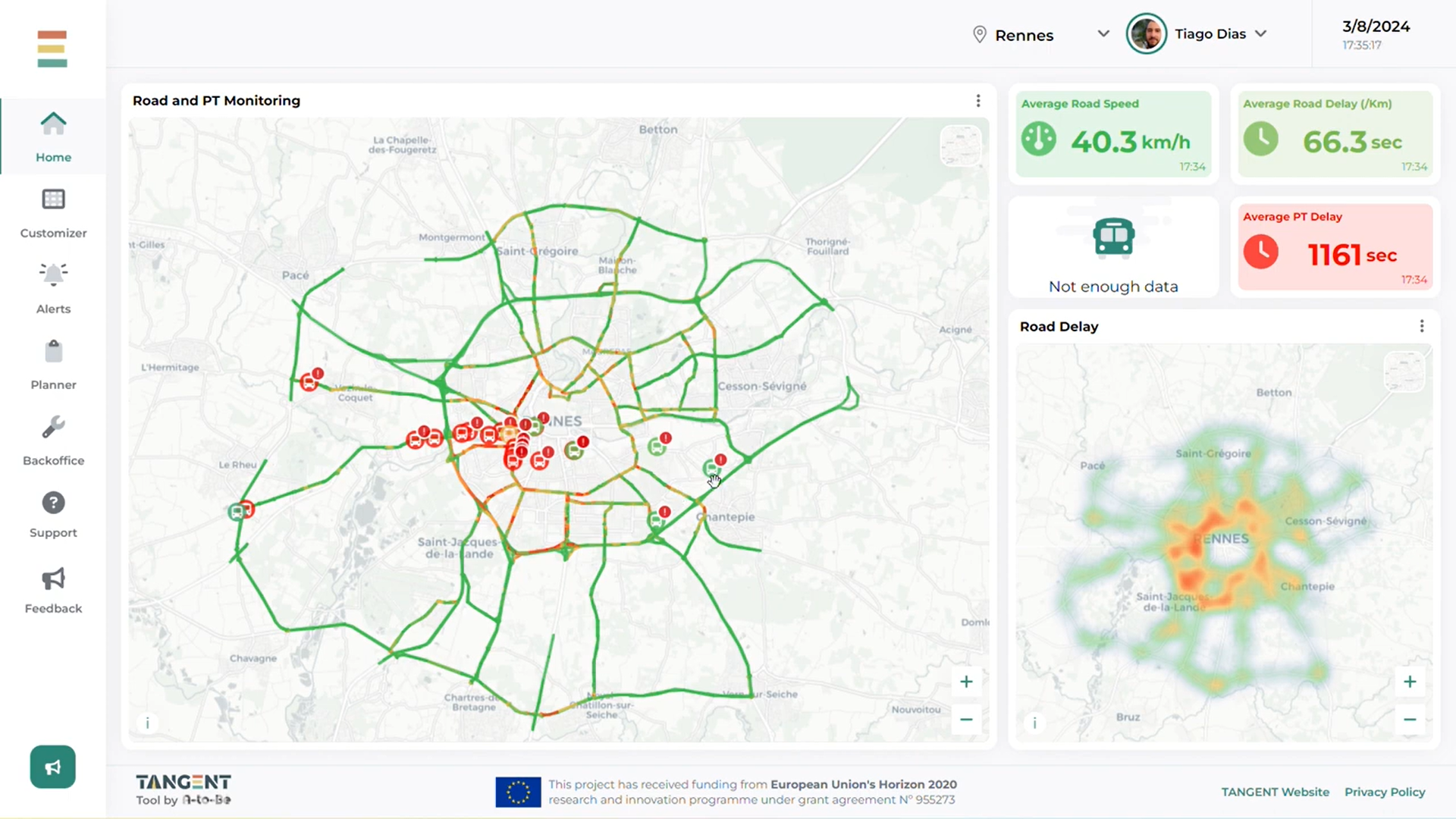}
    \caption{TANGENT Dashboard for Rennes.}% showing complications on the western part of the city.}
   \label{fig:dashboard}
\end{figure}

%\subsubsection{User evaluation} 
\textbf{User evaluation} 
To gather opinions and feedback from real users, we performed a qualitative evaluation involving 43 business stakeholders such as transport operators and authorities.
%Concerning the Data Catalogue, the main insights were related to (i) the importance of providing accompanying guidelines on how to compile metadata fields to gather high-quality metadata, (ii) the need for providing multiple filtering options to quickly navigate the data sources according to the values of different metadata fields (e.g., to combine multiple filters).
%We performed a second development cycle within TANGENT to improve the  Data Catalogue taking into account the feedback received. 
%The final release of the  Data Catalogue was very positively received by industrial partners and transport authorities involved in the testing. 
They highlighted the importance of a Data Catalogue with structured metadata descriptions to reduce the current scattering of information on data sources and facilitate their retrieval. Indeed, we experienced this difficulty ourselves during the data collection phase: retrieving updated and consistent information about data sources often required the involvement of different people within the same company and/or third-party solution vendors. Based on the evaluation feedback, we also improved the Data Catalogue by customising filtering operators to further facilitate data discovery.
In parallel, we performed the testing of the integrated TANGENT solution (cf. Figure~\ref{fig:dashboard}) for the visualization of the current and forecast status of the multimodal network for city and transport authorities, integrating different data-driven services for intelligent incident detection, response plan prescription and the cooperative management of incidents \cite{tra2024dashboard}. The stakeholders involved in its evaluation highlighted the advantage of having data from different transport modes available on the same platform in integrated visualizations. The final detailed assessment of the case studies will be included in the future TANGENT deliverable D7.3-D7.6 (due Q4 2024).

%\item significance evaluation: qui cercherei di spiegare qual è il vantaggio delle tecnologie semantiche rispetto alle alternative (in termini di standardizzazione/interoperabilità, costo e manutenibilità dei mapping, evitando integrazioni punto-punto che sarebbero state certamente più costose) [Quality of the discussion of the benefits and challenges of adopting Semantic Web and Knowledge Graph technologies for solving the addressed problem and/or with respect to alternative approaches.]
%\subsubsection{Significance evaluation} 
\textbf{Significance evaluation} 
The advantages of the proposed solution over potential alternatives are: (i) the collection of structured and actionable metadata in a common machine-readable format that facilitates the findability of heterogeneous data source through the Data Catalogue, and (ii) the reduction of the integration effort for downstream interoperable data-driven services, through the Semantic Harmonisation and Fusion Pipelines and the subsequent Data API: %, due to the hidden complexity that guarantees data interoperability. In the TANGENT solution, 
we removed the need for point-to-point integration between different components and we reduced at minimum the custom development for the different deployments in each urban case study. 

%\item uptake evaluation: qui direi sinceramente che l'uptake corrente è limitato ai soli stakeholder coinvolti nel progetto, che però hanno sottolineato come una soluzione simile manchi completamente nelle loro infrastrutture esistenti e quanto sia importante averla, sia per governance interna (es. catalogo) sia per traffic management (es. dashboard); se ci fosse qualche deliverable di tangent che dà una vista "business", si potrebbe cercare di pescare qualcosa [Proof or plan for large-scale deployment or adoption in the specific domain.]

%\item impact evaluation: qui sottolineerei come la creazione delle data api (basata sul reference model) permetta di abilitare potenzialmente tanti usi e riusi dei dati, "nascondendo" la complessità che sta dietro e come l'approccio sia replicabile in qualsiasi altro dominio [Technological, business, and social impact of the proposed solution, especially in contrast to alternative approaches. + Validity and applicability of the proposed approach in a different domain.]
%\subsubsection{Uptake and impact evaluation} 
\textbf{Uptake and impact evaluation} 
The current uptake of the solution is confined to the involved stakeholders in the four urban case studies, which however represent a significant sample of the target market of traffic management solutions;  an analysis of the possible exploitation in other cities is currently ongoing also by leveraging the services offered by the Horizon Results Booster\footnote{\url{https://www.horizonresultsbooster.eu/}}. Indeed, the involved stakeholders  pointed out the absence of a similar solution in their existing infrastructures and its relevance for both internal governance and traffic management. Concerning the potential impact, we defined a reference conceptual model based on the semantics of existing standards that could be adopted (and possibly extended) to foster interoperability of different solutions for traffic management. Similarly, we demonstrated how to define a metadata extension that supports compatibility with other data portals (leveraging mobilityDCAT-AP) but fulfils additional requirements. Finally, the technological solutions developed for traffic management are not dependent on the mobility domain: with the opportune use of domain ontologies and the definition of specific mappings, the transformation and integration pipelines can be easily configured and adapted to any other domain or market, bringing the same interoperability advantages.

%\item lessons learned: qui forse riprenderei le 5 challenge che hai messo all'inizio e direi cosa abbiamo imparato per ciascuna di esse [Applicability of the lessons learned from the adoption of Semantic Web and Knowledge Graph technologies both from a technical and non-technical perspective.]
%\subsubsection{Lessons learned} 
\textbf{Lessons learned} 
We now discuss some of the lessons learned referencing the five challenges  in Section~\ref{sec:challenges}. Regarding the \emph{Locate} challenge, we experimented the difficulty of obtaining good quality metadata. Structured descriptions of existing data sources are often not available, and the Data Catalogue not only enforced common metadata descriptors but also provided users with useful guidelines to collect high-quality metadata (e.g., guided and dynamic forms for metadata insertion). Concerning the \emph{Access} challenge, we experienced the advantage of having an integrated solution for data findability and access. %The possibility of updating the metadata in the Data Catalogue with a uniform and consistent way to access different data sources or different representations of the same data source makes a huge difference. 
Indeed, often data portals act as simple metadata catalogues, only referencing existing data sources and without taking into account difficulties in accessing the actual data (e.g., authorization, missing documentation of data services, etc.); our Data Catalogue also incorporates and hides the complexity of data harmonisation, giving access to a uniform data API. Considering the \emph{Harmonise} challenge, it is often hard to define common semantics to support different use cases. For this reason, it is important to leverage the semantics already encoded in existing standards without reinventing the wheel and to facilitate the adoption of ontologies by domain experts. Regarding the \emph{Integrate} challenge, it is often difficult to integrate certain types of data (e.g., geographical data considering different location referencing methods or identifiers of stop stations across different transport modes). In these cases, we managed to implement complex transformations by defining mapping rules that integrate custom functions, and by leveraging data fusion with external data sources that specify the correct correspondence between values. Finally, concerning the \emph{Extract} challenge, we demonstrated that a reference ontology supports the generation of harmonised outputs also in formats different from RDF (in our case, CSV for model training and JSON schemas for runtime interactions). This not only facilitates the integration with systems unable to process RDF, but also reduces the size of exchanged data (and, consequently, latency) by avoiding possibly verbose RDF representations. 

\section{Conclusions}\label{sec:conclusions}

This paper has comprehensively described the proposed solution to address data interoperability challenges within a complex scenario of traffic management and its validation within four European cities. The solution guarantees interoperable descriptions of the data sources and applies the any-to-one centralized approach for semantic interoperability enabling data exchange with unambiguous and shared meaning.

The proposed solution consists of four components: the Data Catalogue, for sharing uniform data source descriptions according to the TangentDCAT-AP metadata profile and for enforcing governance mechanisms; the Reference Conceptual Model, a reference ontology defining common semantics for multimodal traffic data; the Semantic Harmonisation and Fusion Pipelines,  to fulfil heterogeneous data integration requirements; and the Data API, a uniform mechanism to access all data sources. %It enables the integration of shared and harmonised data with data-driven services that exploit the data.
Semantic Web technologies proved their efficacy in addressing data interoperability challenges and providing a production-ready to integrate data in downstream services and applications. %, even amidst a complex scenario like multimodal traffic management, which entails diverse stakeholders and data providers.

% The Supplemental Material Statement should be placed at the end of the paper, just before the References (and before Acknowledgements, if present). It counts within the 15-page limit. 
\begin{small}
\paragraph*{Supplemental Material Statement:}
Public deliverables are available on the TANGENT website at \url{https://tangent-h2020.eu/deliverables/}. Implementation reports and related artifacts (e.g., source code) for the TANGENT solution are part of confidential deliverables and can not be shared. The Reference Conceptual Model is published online at (\url{https://github.com/cefriel/tangent-model}), TangentDCAT-AP and controlled vocabularies at (\url{https://github.com/cefriel/tandcatap}). The Chimera framework used for the implementation of the pipelines is available at (\url{https://github.com/cefriel/chimera}).

\subsubsection*{Acknowledgments}
The presented research was partially supported by the TANGENT project (Grant Agreement 955273) funded by the European Commission under the Horizon 2020 Research and Innovation Programme.

This preprint has not undergone peer review or any post-submission improvements or corrections. The Version of Record of this contribution will be published in The Semantic Web – ISWC 2024.
%is published in [insert volume title], and is available online at https://doi.org/[insert DOI]
\end{small}

\bibliographystyle{splncs04}
\bibliography{biblio}

\end{document}